\documentclass[aps,prl,twocolumn,showpacs,groupedaddress]{revtex4}  
\usepackage{amsmath}
\usepackage{epsfig}
\usepackage{setspace}

\usepackage{graphicx}  
\usepackage{dcolumn}   
\usepackage{bm}        
\usepackage{amssymb}   

\begin{document}



\title{Gaussian flexibility with Fourier accuracy: the periodic von
Neumann basis set}
\author{Asaf Shimshovitz and David J.~Tannor}                             
\affiliation{Department of Chemical Physics, Weizmann Institute of Science, Rehovot, 76100 Israel}
\date{\today}

\begin{abstract}
We propose a new method for solving quantum mechanical problems,
which combines the
flexibility of Gaussian basis set methods with the numerical
accuracy of the Fourier method.  The method is based on the
incorporation of periodic boundary conditions into the von Neumann
basis of phase space Gaussians [F. Dimler et al., New J. Phys. 11, 105052 (2009)]. 
In this paper we focus on the Time-independent Schr\"odinger Equation 
and show results for the harmonic, Morse and Coulomb potentials
that demonstrate that the periodic von Neumann method or pvN is significantly more
accurate than the usual vN method. Formally, we are able to show an
exact equivalence between the pvN and the Fourier Grid Hamiltonian
(FGH) methods. Moreover, due to the locality of the pvN functions we
are able to remove
Gaussian basis
functions without loss of accuracy, and obtain significantly better   
efficiency than that of the FGH. We show
that in the classical limit the method has the remarkable
efficiency of 1 basis function per 1 eigenstate. 
\end{abstract}

\maketitle

The formal framework for quantum mechanics is an infinite dimensional
Hilbert space. In any numerical calculation, however, a wave function
is represented in a finite dimensional basis set and therefore the 
choice of basis set determines the accuracy. The optimal basis set should
combine accuracy and flexibility, allowing a small number of basis
functions to represent the wave functions even in the presence of
complex boundary conditions and geometry. Unfortunately, these two
criteria ---accuracy and efficiency--- are usually in conflict, and globally accurate methods
\cite{fourier_method,fgh,wei}
lack the flexibility of local methods
\cite{dgb,garashchuk}.     
For example, in the 
pseudospectral Fourier grid method the wave function is represented by its values on a finite number of evenly spaced grid points. Due to the Nyquist sampling theorem, this allows for an exact representation of the wave provided the wavefunction is band limited with finite support\cite{whittaker,nyquist,shannon}. However, the non-local form of the basis functions in momentum space leads to limited efficiency. 
On the other hand, in the von Neumann basis set \cite{brixner,von_neumann} each basis function is
localized on a unit cell of size $h$ in phase space. However, despite the formal completeness
of the vN basis set\cite{perelomov}, attempts to utilize
this basis in quantum numerical calculations 
have been plagued with numerical errors\cite{davis_and_heller,poirier}.

The purpose of this paper is to establish a precise mathematical formalism for the von Neumann basis on a truncated phase space. This allows us to demonstrate an exact equivalence with the Fourier method while retaining the flexibility of a Gaussian basis. It puts on a rigorous basis the seminal work of Dimler et al.\cite{dimler} who used a vN basis with periodic boundary conditions, although in our formalism the periodicity of the vN basis appears only implicitly.



The von Neumann basis set \cite{von_neumann} is a subset of
the ``coherent states'' of the form:
\begin{eqnarray}
g_{nl}(x)=\left(\frac{2\alpha}{\pi}\right)^\frac{1}{4}\exp\left(-\alpha(x-na)^{2}-il\dfrac{2\pi\hbar}{a}(x-na)\right)
\end{eqnarray}%
where $n$ and $l$ are integers. Each basis function is a Gaussian centered
at $(na,\frac{2\pi l}{a})$ in phase space. 
The parameter
$\alpha=\frac{\sigma_p}{2\sigma_x}$ controls the FWHM of each Gaussian in $x$ and
$p$ space. Taking $\Delta x =a, \Delta p
=h/a$ as
the spacing between neighboring Gaussians in $x$ and $p$ space
respectively, we note that $\Delta x \Delta p=h$
so we have exactly one basis function per unit cell in
phase space. As shown in\cite{perelomov} this  implies completeness in the Hilbert space.

The ``complete'' vN basis, where $n$ and $l$ run over all integers, spans the
infinite Hilbert space. In any numerical calculation, however, $n$ and
$l$ take on a finite number of values, 
producing $N$ Gaussian basis functions $\lbrace g_i(x)\rbrace$, $i=1...N$. 
Since the size of one vN unit cell is
$h$, the area of the truncated vN lattice is given by   
$S^{\rm vN}=Nh$.  

\begin{figure}[h]
\begin{center}
\includegraphics [width=4cm]{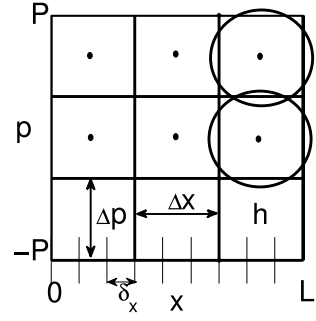}
\end{center}
\begin{center}
\caption{\footnotesize{$N=9$ coordinate grid points and $N=9$ vN unit cells span
the same area in phase space,$S=Nh$. The vN basis
functions are Gaussians located at the center of each unit
cell. }}
\label{VN}
\end{center}
\end{figure}

In the pseudospectral Fourier method, a function $\psi(x)$ that is
periodic in $L$ and band limited in $K=\frac{P}{\hbar}$ can be written
in
the following form: 
$\psi(x)=\sum_{n=1}^{N} \psi(x_n)\theta_{n}(x)$,
where $x_n=\delta_x(n-1)$, and
$\delta_x=\frac{\pi\hslash}{P}=\frac{L}{N}$. 
The basis functions $\lbrace\theta_n(x) \rbrace$ are given by 
\cite{tannor_book}:
\begin{eqnarray}
\theta_n(x)=\sum_{j = \frac{-N}{2} + 1}^{\frac{N}{2}}\frac{1}{\sqrt{LN}}\exp\left(\frac{i2\pi j}{L}(x-x_n)\right),
\end{eqnarray}
which can be shown to be sinc functions that are periodic on the domain $[0,L]$.
The set $\lbrace\theta_i(x) \rbrace$ $i=1,..,N$ spans a
rectangular shape in phase space with area of $S^{\rm
FGH}=2LP=2L\frac{\pi\hbar}{\delta_x}=Nh$.
Thus $N$ unit cells in the vN lattice and $N$ grid points in the Fourier method cover the same rectangle with an area in phase space of: 
\begin{eqnarray}
S^{\rm vN}=S^{\rm FGH}=Nh
\label{S1}
\end{eqnarray}
(Fig. \ref{VN}). This suggests that $N$ vN basis functions confined to
this area will be equivalent to the Fourier basis set.  
Unfortunately, the attempt to use $N$ Gaussians as a basis set for the
area in eq.(\ref{S1}) (Fig. \ref{VN}) is unsuccessful, a consequence of the
Gaussians on the edges protruding from the truncated space.
However, by combining the Gaussian and the Fourier basis functions we
can generate a ``Gaussian-like'' basis set that is confined to the
truncated space. We use the basis sets $\lbrace g_i(x) \rbrace$ and
$\lbrace\theta_i(x) \rbrace$ to construct a new basis set, $\lbrace
\Tilde g_i(x) \rbrace$: 
\begin{eqnarray}
\Tilde g_m(x)=\sum_{n =  1}^{N} \theta_n(x)g_m(x_n)
\label{dvn}
\end{eqnarray}
for $m=1,...,N$. The new basis set is in some sense, the Gaussian functions with periodic boundary conditions. We can write eq.(\ref{dvn}) in matrix notation as: $\Tilde G=\Theta G$
where $G_{ij}=g_j(x_i)$ 
By taking the width parameter $\alpha=\frac{\Delta p}{2\Delta x}$ we
can guarantee that the pvN functions have no linear dependence and
that the matrix $G$ is invertible, that is $\Tilde G G^{-1}=\Theta$.
The invertibility of $G$ implies that both bases span the same space. 

The representation of $\vert \psi \rangle$ in the pvN basis set is given by:
\begin{eqnarray}
\vert \psi \rangle=\sum_{m =  1}^{N} \vert\Tilde g_m\rangle a_m.
\label{psi}
\end{eqnarray}
To find the coefficients $a_m$ we first define the overlap matrix, $S$: 
\begin{eqnarray}
S_{ij}&=& \langle\Tilde g_i \vert \Tilde
g_j\rangle=\int_{0}^{L} \Tilde g_i^{*}(x)\Tilde g_j(x)dx
\nonumber \\
&=&\sum_{n =  1}^{N}\sum_{m =  1}^{N} g_i^{*}(x_n)
g_j(x_m)\int_{0}^{L} \theta_n^{*}(x)\theta_m(x)dx
\nonumber \\
&=&\sum_{n =  1}^{N} g_i^{*}(x_n)g_j(x_n)
\label{ip}
\end{eqnarray}
or
\begin{equation}
S=G^{\dag}G.
\label{S}
\end{equation}
Using the completeness relationship for non-orthogonal bases, 
$\vert \psi \rangle$ can be expressed as
\begin{equation}
\vert \psi \rangle=\sum_{n =  1}^{N} \sum_{m =  1}^{N}
\vert\Tilde g_m \rangle (S^{-1})_{mn} \langle\Tilde
g_n\vert\psi\rangle.
\label{psi2}
\end{equation}
Comparing with eq.(\ref{psi}) we find that $a_m=\sum_{n =  1}^{N} (S^{-1})_{mn} \langle\Tilde g_n\vert\psi \rangle$
and $\langle\Tilde g_i\vert\psi \rangle=\sum_{n =  1}^{N} g_i^{*}(x_n)\psi(x_n)$.

Although the periodic von Neumann (pvN) and the Fourier methods span the same rectangle in phase
space, the localized nature of the basis functions in the pvN method can lead to significant advantages. In particular, if $\vert\psi\rangle$ has an irregular phase space shape we may expect that some of the pvN basis functions will fulfill the relation: $\langle\Tilde g_j|\psi\rangle=0$, $j=1,...,M$. Due to the non-orthogality of the basis we cannot simply eliminate the states $\tilde{g}_j$, since the \textit{coefficients} of $\tilde{g}_j$ may include contributions from remote basis functions, but 
we can take advantage of the vanishing overlaps by defining a bi-orthogonal von Neumann basis (bvN) $\{{b}_i(x)\}$. 
\begin{eqnarray}
\vert b_i\rangle=\sum_{j =  1}^{N} \vert\Tilde g_j\rangle (S^{-1})_{ji}
\label{fvn}
\end{eqnarray}
or in matrix notation: $B=\Tilde GS^{-1}.$ 
Inserting eq.\ref{fvn} into eq.\ref{psi2}, $\vert \psi \rangle$ can be
written as 
\begin{eqnarray}
\vert \psi \rangle=\sum_{n =  1}^{N} \vert b_n\rangle c_n=\sum_{n =  1}^{N} \vert b_n\rangle \langle \Tilde g_n\vert\psi\rangle. 
\end{eqnarray}
By assumption, $M$ of the coefficients are zero, hence in order to
represent $|\psi\rangle$ in the bvN basis set  we need only $N'=N-M$
basis functions.
Note that the bvN and pvN are bi-orthogonal bases, meaning that each
set taken by itself is non-orthogonal but they are orthogonal to each other. This
can be shown easily by:
\begin{eqnarray}
\langle \Tilde g_i\vert b_j\rangle&=&\sum_{n =  1}^{N} g^{*}_i(x_n)f_j(x_n)
\nonumber\\
&=&\sum_{m =  1}^{N} \sum_{n =  1}^{N} g^{*}_i(x_n)g_m(x_n)(S^{-1})_{mj}
\nonumber\\
&=&\sum_{m =  1}^{N} S_{im}(S^{-1})_{mj}=\delta_{ij}.
\end{eqnarray}
For many practical applications the full knowledge of
the basis wavefunctions is unnecessary: we need only the value of
the basis functions at the sampling points. For example the evaluation
of Hamiltonian matrix elements can be performed explicitly by:

\begin{eqnarray}
H^{\rm pvN}_{ij}&=&\langle\Tilde g_i\vert H\vert \Tilde g_j\rangle
\nonumber\\
&=&\sum_{m =  1}^{N}\sum_{n =  1}^{N} g_i^{*}(x_m)\langle\theta_m\vert H\vert\theta_n\rangle g_j(x_n)
\nonumber\\
&=& \sum_{m =  1}^{N}\sum_{n =  1}^{N} g_i^{*}(x_m)H^{\rm FGH}_{mn}g_j(x_n) 
\end{eqnarray}
and similarly:
\begin{eqnarray}
H^{\rm bvN}_{ij}=\sum_{m =  1}^{N}\sum_{n =  1}^{N}
b_i^{*}(x_m)H^{\rm FGH}_{mn}b_j(x_n) 
\end{eqnarray}
where $H^{\rm FGH}=V^{\rm FGH}+T^{\rm FGH}$ and the potential and the
kinetic matrix are given by: $V^{\rm FGH}_{ij} \approx V(x_i)\delta _{ij}$
and
\begin{eqnarray}
T^{\rm FGH}_{ij} =\frac{\hslash^{2}}{2M}
\begin{cases}
  \frac{K^{2}}{3}(1+\frac{2}{N^{2}}),  & \mbox{if} \quad i = j \\
  \frac{2K^{2}}{N^{2}}\frac{(-1)^{j-i}}{\rm {sin}^{2}(\pi\frac{j-i}{N})}, & \mbox{if} \quad i \neq j
\end{cases}
\end{eqnarray}
\cite{tannor_book2}. The eigenvalue problem in a non-orthogonal basis set becomes 
$HU=sUE$; in the pvN basis set  $s$ is given by 
eq. (\ref{S}) and in the bvN basis set $s$ is given by: 
\begin{equation}
B^{\dag}B=S^{-1}G^{\dag}GS^{-1}=S^{-1}.
\end{equation} 
Diagonalization should give accurate results for all wavefunctions
localized to the classically allowed region of the rectangle. 

\begin{figure}[h]
\begin{center}
\includegraphics [width=8cm]{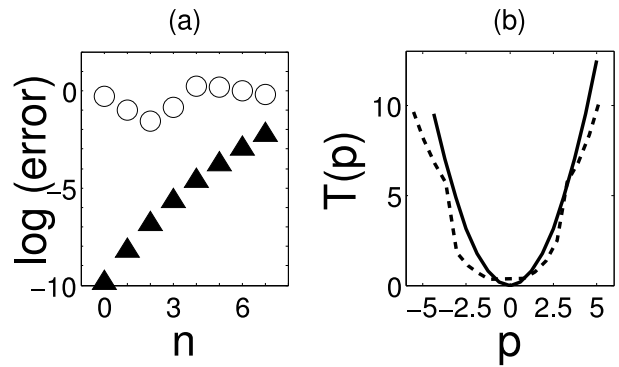}
\end{center}
\begin{center}
\caption{\footnotesize{ (a) Error in the lowest 8 eigenvalues of the harmonic oscillator. vN({\large$\circ$}), pvN($\blacktriangle$). (b) Kinetic energy spectra using 16 basis functions. vN(dashed), pvN(solid).}}
\label{harmonic}
\end{center}
\end{figure}

As a numerical test of the pvN basis set we studied the standard example of the
harmonic oscillator $V(x)=\frac{m\omega^{2}x^{2}}{2}$ in units such
that $m=\hbar=\omega=1$. We calculated the first 8
eigenenergies using 16 pvN compared with 16 conventional Gaussian basis functions.
In the Gaussian basis set the Hamiltonian and the overlap  matrices
were calculated analytically as: $H_{ij}=\langle
g_i|H|g_j\rangle=\int_{-\infty}^{\infty}g^{*}_i(x)[-\frac{d^{2}}{dx^{
2 }}+V(x)]
g_j(x)dx$
and
$S_{ij}=\langle
g_i|g_j\rangle=\int_{-\infty}^{\infty}g^{*}_i(x)g_j(x)dx$. 
The 16 sampling points were taken from -5 to 5-$\delta_x$ and the
width parameter was $\alpha=0.5$. The results, shown in Fig.
\ref{harmonic}, show the superiority of the pvN basis set compared to
the standard Gaussian basis set. In fact, the results obtained with
the pvN basis set are exactly as accurate as in the Fourier grid
method. The kinetic energy spectra in Fig. \ref{harmonic} reveal
the deficiency of the conventional Gaussian scheme.

\begin{figure}[h]
\begin{center}
\includegraphics [width=8cm]{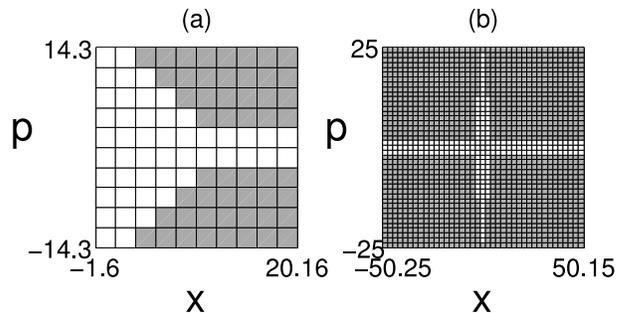}
\end{center}
\begin{center}
\caption{\footnotesize{Phase space area spanned in
the bvN method (white) and in the pvN (or FGH) method (full rectangle) for
Morse (a) and Coulomb (b).}}
\label{phase}
\end{center}
\end{figure}

In the bvN basis set we are able to remove some of the basis functions
and construct lower dimensional $H^{\rm bvN}$ and $S^{\rm bvN}$
matrices without losing accuracy.  In order to test this claim, we
calculated numerically the eigenenergies of the Morse oscillator
$V(x)=D(1-e^{-\beta x})^{2}$ and the Coulomb potential
$V(x)=-\frac{Q^{2}}{x}$  by using both the FGH and fvN basis sets.
The
Morse parameters were taken to be $D=12$, $m=6$, $\beta=0.5$ and
$\hbar=1$. For FGH, 100  grid points between $[-1.6,20.1]$ were
required in order to get 4 digits of accuracy in energy for all 24
bound states. By using the bvN basis functions (constructed from
10$\times$10 vN functions with $\alpha=0.5$)
we obtain the same 4 digit accuracy with only 48 basis functions.
This is demonstrated graphically in Fig. \ref{phase} (a). The
figure shows the phase space representation of 100 evenly grid points.
Although it requires 100 pvN basis functions to span this area in phase space,
due to the flexibility of the bvN basis set we can 
suffice with just the basis functions in the  classically  allowed
region (white squares). 

\begin{figure}[h]
\begin{center}
\includegraphics [width=6cm]{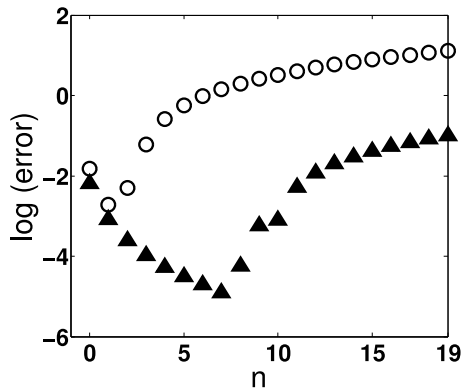}
\end{center}
\begin{center}
\caption{\footnotesize{The deviation between the calculated and the
exact eigenenergies for the lowest 20 eigenstates of the Coulomb
potential using 190 basis functions for the bvN ($\blacktriangle$) and
pvN (or FGH)({\large$\circ$})}}
\label{coulomb}
\end{center}
\end{figure}

In the Coulomb potential the efficiency of the bvN basis set is even higher. The rectangular shape in phase space of the FGH is a very wasteful representation for the cross-like shape of the eigenstates. 
The Coulomb parameters were taken to be $Q=1$, $m=1$ and $\hbar=0.5$.
For FGH 1599 grid points between $[-50.2, 50.1]$  were required
in order to get 4 digits of accuracy in energy for the first 9 excited
states. By using the bvN basis functions (constructed from
39$\times$41 vN functions with $\alpha=0.2367$) only in the
classically allowed energy shell we obtain the same accuracy with only
189 basis functions. Fig. \ref{phase} (b)illustrates the
efficiency of the bvN basis by showing the phase space area in FGH and
bvN bases. Figure \ref{coulomb} shows the error in the eigenenergies by
using 190 functions in the FGH and in the bvN method
respectively.

\vspace{-0.5cm}

\begin{figure}[h]
\begin{center}
\includegraphics [width=6cm]{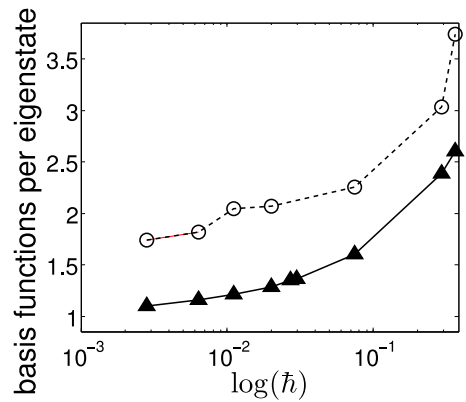}
\end{center}
\begin{center}
\caption{\footnotesize{Basis efficiency for the Morse potential as
function of $\hbar$. For pvN (or FGH) (dashed) and bvN(solid)}}
\label{ratio}
\end{center}
\end{figure}

The ability to localize a bvN function at a specific point in phase
space results in the remarkable concept of 1 basis function per 1
eigenstate. This means that in order to calculate $N$ eigenenergies we
need only $N$ basis functions. Obviously, such one per one efficiency,
if reachable, will be the ideal efficiency for any basis set.  In order to test the ability of the bvN method to reach the
ideal efficiency we examined the Morse potential and
looked for the smallest number of bases that will provide exact values
of the energies (12 digits of accuracy) for all the eigenstates in the
energy shell $E=11.25$. The bvN method indeed, tends to the ideal
efficiency in the classical limit where $\hbar\rightarrow 0$
(Fig. \ref{ratio}). This remarkable result is unique for
the bvN method. For example, the efficiency of the Fourier method has
an upper bound, which can be calculated analytically as $\eta=1.6959$
(the ratio of the classically allowed phase space to the area of the
rectangle).




In summary, we introduced the ``Gaussian-like'' pvN basis set and showed
the equivalence between the pvN and the Fourier method.
Several TISE problems were tested. We showed that the pvN method is
as accurate as the Fourier method; however, due to the localization
of the basis functions, we can construct a flexible basis set, bvN,
which is much more efficient then the FGH.
We also have shown that in the classical limit the number of basis
functions is equal to the number of the exact eigenvalues. This
remarkable ``1 per 1'' quality, is of course, an optimal result for
any basis set, and as far as we are aware has never been achieved
before.

As discussed above, the use of primitive Gaussian functions on the vN lattice yields disappointing results
\cite{davis_and_heller,poirier}. The success of our method
is due to two key points:
1. By defining the basis set as a linear combination of the   $\{\theta_i(x)\}$ we ensure that the basis indeed spans a band limited region
with finite support.
2. By using the bvN basis set the coefficients become locallized. In the conventional Gaussian basis set the delocalized coefficients arise from the non-locality of $S^{-1}$.

In this paper we focused on model 1-d TISE problems. However, Gaussians 
are used as basis functions in a variety of methods
for solving the TISE and
TDSE
\cite{hk,shalashilin,fg,martinez}.
All these
methods have suffered from numerical difficulties resulting from the
non-locality of $S^{-1}$ and we believe that significant improvement
is possible using the pvN and the bvN basis sets.
Finally, the pvN method is not limited to quantum mechanical problems.
There is a large literature on the use of a Gaussian basis set in signal
processing where it goes by the name of the Gabor transform
\cite{gabor,bastians,ingrid,orr,porat}.   The
Gabor transform is known to have problems with stability, which can be traced 
to the ``no-go" theorem of Balian and
Low\cite{balian,low}---the statement that localized basis sets are incompatible with
orthogonal basis sets.  While the pvN and bvN bases do not violate the
no-go theorem, they seem to effectively circumvent it.  We therefore
believe they can have  a significant impact on signal processing in
general.

This work was supported by the Israel Science Foundation and made possible in part by the historic generosity of the Harold Perlman family.






\begin{thebibliography}{99}



\bibitem{fourier_method}
R.Kosloff in \textit{Numerical Grid Methods and their Application to
Schr\"odinger's Equation} ed. C. Cerjan (Kluwer, Boston, 1993)
\bibitem{fgh}
C. C. Marston and G. G. Balint-Kurti, J.Chem.Phys. {\bf 6}, 3571
(1989).
 \bibitem{wei}
G. W. Wei, J.Phys.B. {\bf 33}, 343 (2000)
\bibitem{dgb}
Z.Ba\v{c}i\'{c}, R.M Whitnell, D.Brown and J.C.Light, Comp. Phys. Comm. {\bf
51}, 35 (1988)
\bibitem{garashchuk}
S. Garashchuk and J. C.Light, J.Chem. Phys. {\bf 114}, 3929
(2001)
\bibitem{whittaker}
E. T. Whittaker, Proc. R. Soc. Edinburgh {\bf 35}, 181 (1915).
\bibitem{nyquist}
H. Nyquist, Trans. AIEE 1 {\bf 47}, 617 (1928).
\bibitem{shannon}
C. E. Shannon, Proc. IRE {\bf 37}, 10 (1949).
\bibitem{brixner}
S. Fechner, F. Dimler, T. Brixner, G. Gerber and D.
J.Tannor, Opt. Express {\bf 15}, 15389 (2007)
\bibitem{von_neumann}
J.von Neumann, Math. Ann. {\bf 104}, 570 (1931)
\bibitem{perelomov}
A.M Perelomov, Theor. Math.Phys {\bf 11}, 156 (1971)
\bibitem{davis_and_heller}
M.J.Davis and E J.Heller, J.Chem. Phys. {\bf 71}, 3383 (1979)
\bibitem{poirier}
B. Poirier and A.Salam, J.Chem. Phys. {\bf 121}, 1690 (2004)
\bibitem{dimler}
F. Dimler, S. Fechner, A. Rodenberg,  T. Brixner, and D.
J.Tannor, New J. of Phys. {\bf 11}, 105052 (2009)
\bibitem{tannor_book}
D. J. Tannor, \textit{Introduction to Quantum Mechanics: A Time-dependent
Perspective} (University Science Books, Sausalito, 2007), eq.11.163.
\bibitem{tannor_book2}
Ref.\cite{tannor_book} eq.11.172.
\bibitem{mapped_fourier}
E. Fattal, R. Baer and R. Kosloff, Phys. Rev. E. {\bf 53}, 1217
(1996)
\bibitem{hk}
M. F. Herman and E. Kluk, J.Chem. Phys. {\bf 91}, 27 (1984)
\bibitem{shalashilin}
D. V. Shalashilin
and M. S. Child, J.Chem. Phys. {\bf 113}, 10028 (2000)
\bibitem{fg}
E J.Heller, J.Chem. Phys. {\bf 75}, 2923 (1981) 
\bibitem{martinez}
M. Ben-Nun and T. J.Martinez, J.Chem. Phys. {\bf 108}, 7244 (1998)
\bibitem{gabor}
D.Gabor, J. Inst. Elect. Eng. {\bf 93}, 429 (1946)
\bibitem{bastians}
M.J.Bastiaans, IEEE {\bf 68}, 538 (1980)
\bibitem{ingrid}
I. Daubechies, IEEE {\bf 36}, 961 (1990)
\bibitem{orr}
R. S.Orr, Signal Processing {\bf 34}, 85 (1993)
\bibitem{porat}
T. Genossar and M. Porat, IEEE {\bf 22}, 449 (1992)
\bibitem{balian}
R. Balian, C. R. Acad. Sci. III {\bf 292}, 1357 (1981)
\bibitem{low}
F. Low, in \textit{A Passion for Physics--Essays in Honor of Geoffrey Chew} (World Scientific, Singapore, 1985), pp.17-22
\end{thebibliography}
\end{document}